\documentclass[runningheads]{llncs}

\usepackage{graphicx}
\usepackage[latin1]{inputenc}
\usepackage{tikz}
\usetikzlibrary{shapes,arrows}
\usepackage{enumerate}
\usepackage{makecell}
\usepackage{cite}
\usepackage{bm}

\begin{document}

\title{ Brain tumour segmentation using cascaded 3D densely-connected U-net }

\author{
Mina Ghaffari\inst{1} \and Arcot Sowmya\inst{2} \and Ruth Oliver\inst{1}
}

\authorrunning{M. Ghaffari et al.}

\institute{Macquarie University, Sydney, Australia \and
University of New South Wales, Sydney, Australia\\
\email{mina.ghaffari@hdr.mq.edu.au}
}

\maketitle              % typeset the header of the contribution

\begin{abstract}
Accurate brain tumour segmentation is a crucial step towards improving disease diagnosis and proper treatment planning. In this paper, we propose a deep-learning based method to segment a brain tumour into its subregions: whole tumour, tumour core and enhancing tumour. The proposed architecture is a 3D convolutional neural network based on a variant of the U-Net architecture of Ronneberger et al. \cite{Ronneberger2015} with three main modifications: (i) a heavy encoder, light decoder structure using residual blocks (ii) employment of dense blocks instead of skip connections, and (iii) utilization of self-ensembling in the decoder part of the network. The network was trained and tested using two different approaches: a multitask framework to segment all tumour subregions at the same time, and a three-stage cascaded framework to segment one subregion at a time. An ensemble of the results from both frameworks was also computed. To address the class imbalance issue, appropriate patch extraction was employed in a pre-processing step. Connected component analysis was utilized in the post-processing step to reduce the false positive predictions. Experimental results on the BraTS20 validation dataset demonstrates that the proposed model achieved  average Dice Scores of 0.90, 0.82, and 0.78 for whole tumour, tumour core and enhancing tumour respectively.

\end{abstract}

\section{Introduction}
Accurate and reliable brain tumour segmentation from neuroimaging scans is a critical step towards improving disease diagnosis and proper treatment planning that increases the survival chance of patients. Brain tumours are highly heterogeneous in terms of shape, size and location, which makes their segmentation challenging. In addition, brain tumours can be highly infiltrating  and it may be difficult to distinguish healthy brain tissue from the tumour. Manual segmentation of brain tumours in MR images is a laborious task that is both time-consuming and subject to rater variability. Therefore, reliable automatic segmentation of brain tumours has attracted considerable attention over the past two decades. Most recent automatic segmentation methods build on convolutional neural networks (CNNs) trained on manually annotated dataset of a large cohort of patients. The Brain Tumour Segmentation (BraTS) challenge  public dataset has become the benchmark in this area \cite{Menze2015, BraTS, Bakas, Bakas2017,Bakas20171, Bakas20172} since 2012. 

The BraTS challenge organizers provide magnetic resonance imaging (MRI) scans in four modalities: T1-weighted (T1), T1-weighted post-contrast (T1c), T2-weighted (T2), and fluid-attenuated inversion recovery (FLAIR) MRIs, with  corresponding manual segmentation. The participants are required to produce segmentation masks of three glioma sub-regions that contain enhancing tumour (ET), tumour core (TC) and whole tumour (WT). The BraTS2020  dataset consists of 369 and 125 cases for training and validation respectively. 

Since the revolution of deep learning and more specifically CNN, the most successful models for BraTS are based on CNN \cite{Ghaffari2020}. The top ranked models submitted to recent BraTs challenges were DeepMedic, that is based on multi-scale processing \cite{Kamnitsas2016}, cascaded Fully  Convolutional  Networks, based on hierarchical binary segmentation \cite{Wang2018} and U-net based models \cite{Isensee2018},\cite{Crimi2019}. 

U-net, which was first introduced in 2015 \cite{Ronneberger2015}, is a CNN architecture consisting of an encoder and a decoder. Due to its straightforward architecture as well as its high segmentation accuracy, U-net or its variants such as V-net \cite{Milletari2016a} are used in most state-of-the-art medical image segmentation tasks \cite{Isensee}, \cite{Li2018}, \cite{Chen2018a} and it has been argued that that 'a well-trained U-net is hard to beat' \cite{Isensee2018}. Hence, while using U-net as the backbone of their network, most BraTS18 and BraTS19 participants focussed on enhancing their model performance by optimizing the preprocessing step, training procedure, co training using local datasets, or applying ensemble learning. Most of the top-ranked models in recent BraTS challenges were inspired by U-net \cite{Ghaffari2020} including all top ranked models of BraTS19. As an example, the top-ranked model in BraTS 2019 was a two stage Cascaded U-net in which the first stage U-net predicts a coarse segmentation map and the second stage U-net provides a more accurate segmentation \cite{Crimi2019}. Zhao et al\cite{Crimi2019} applied a bag of tricks including data sampling, random patch-size training, semi-supervised learning, self-ensembling, result fusion, and warming-up learning rate to enhance their U-net performance. McKinley et al \cite{Crimi2019} modified their 3rd place entry to the BRaTS18: 'DeepSCAN' model, (a shallow U-net-style network with densely connected blocks of dilated convolutions and label-uncertainty loss) by adding a lightweight local attention mechanism and were ranked 3rd again in BraTS19.\\
In this work a modified 3D version of the U-net is utilized, and in order to enhance the model accuracy, pre- and post- processing steps as well as training procedure optimization are applied, and ensembling and data augmentation are utilized during inference time. We report preliminary results on the validation dataset of BraTS20 dataset. The results are computed online using the CBICA Image Processing Portal (\url{https://ipp.cbica.upenn.edu}). The rest of this paper is organized as follows: Methodology including the dataset preprocessing, network architecture, and post-processing are explained in section \ref{methodology}. Experiments and results are provided in section \ref{experimentsAndResults}, followed by a discussion and conclusion in section \ref{conclusion}.

\section{Methodology}\label{methodology}
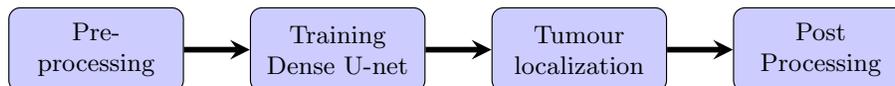
\begin{figure}
    \tikzstyle{block} = [rectangle, draw, fill=blue!20, text width=6em, text centered, rounded corners, minimum height=3.2em, font=\tiny\footnotesize]

\begin{center}
    \resizebox{\columnwidth}{!}{
        \begin{tikzpicture}[node distance = 3cm, auto]
            % Place nodes
            \node [block] (preprocessing) {Pre-processing};
            \node [block, right of=preprocessing] (training) {Training Dense U-net };
            \node [block, right of=training] (localization) {Tumour localization};
            \node [block, right of=localization] (postProcessing) {Post\linebreak Processing};
            % Draw edges
            \draw [line width=2, ->, >=stealth] (preprocessing) -- (training);
            \draw [line width=2, ->, >=stealth] (training) -- (localization);
            \draw [line width=2, ->, >=stealth] (localization) -- (postProcessing);
        \end{tikzpicture}
    }
\end{center}
    \caption{Top-level schematic of the proposed model} \label{model}
\end{figure}

A top level diagram of the proposed method is illustrated in Fig.~\ref{model}. The proposed model is composed of four different modules. In this section, each of these modules is discussed in more detail.

\subsection{Dataset Preprocessing}
The BraTS20 dataset consists of 369 multi-institutional pre-operative multimodal MRI scans. This dataset already has been put through various pre- processing steps by the organizers, so that all images are skull-stripped, have isotropic resolution, and are co-registered MR volumes \cite{Menze2015}. MR scans often contain intensity non-uniformities due to magnetic field inhomogeneity, which can impact the segmentation results.  To compensate for this, a bias field correction algorithm was applied using N4ITK \cite{Tustison2010}.  All modalities were  then  normalized  by  subtracting  the  mean  from  each  of  their  voxels  and dividing by the standard deviation of the intensities within the brain region of that image, so that each modality has zero mean and unit variance. In order to address the small size of the dataset, data augmentation was performed using random rotation (-6 and 6 degree),  scaling (0.9..1.1) and mirroring. 

\subsection{Network Architecture}
\begin{figure}[htp]
    \includegraphics[scale=0.50]{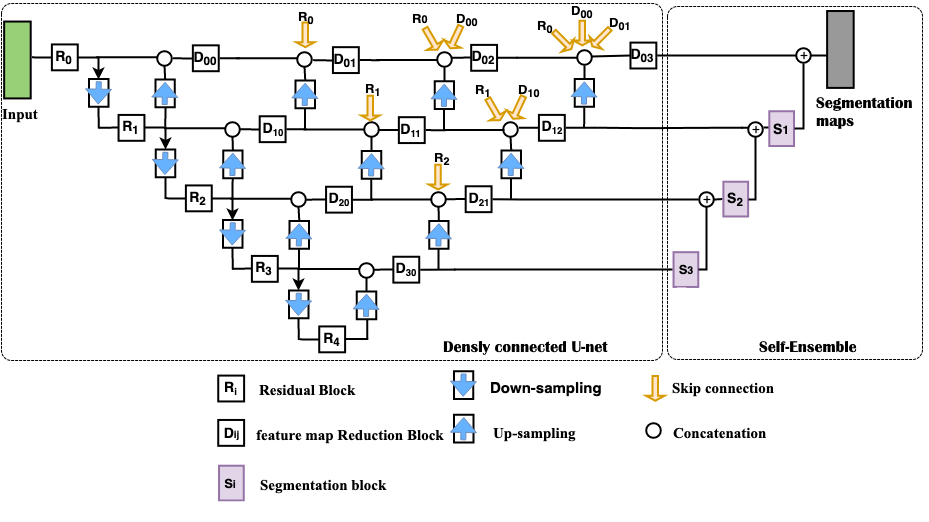}
    \centering
    \caption{Schematic visualization of the network architecture \cite{Ghaffari2020b}} \label{Proposed_network}
\end{figure}

Inspired by other work \cite{Zhou2018}, we propose a modified 3D version of the well known U-net, as shown in Fig.~\ref{Proposed_network}. The main differences between this network and a generic U-net are threefold:  (i) a heavy encoder, light decoder structure using residual blocks \cite{He2016} (ii) employment of dense blocks instead of skip connections \cite{Zhou2018} and (iii) utilization of self-ensembling in the decoder part of the network.

\begin{figure}[htb!]
    \includegraphics[scale=0.35]{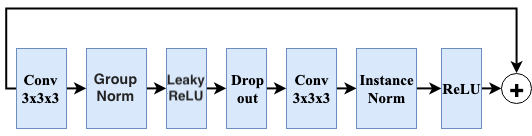} 
    \centering
    \caption{Schematic illustration of the residual block used in the proposed Network \cite{Ghaffari2020b} } \label{Residual_Blocks}
\end{figure} 

\begin{enumerate}[(i)]
\item The residual block as shown in Fig.~\ref{Residual_Blocks} consists of two $3\times3\times3$ convolutions and Group Normalization \cite{Wu2020} with group size of 8 and Rectified Linear Unit (ReLU) activation, followed by additive identity skip connection. Residual blocks learn a non-linear residual that is added to the input and provide a deeper architecture to improve the gradient flow.
\item Dense convolutional blocks merge the feature maps at each level of the network by integrating the output of all previous convolutional blocks in the same level with the upsampled output of the corresponding lower-level dense block. This method provides the network decoder with more semantic information from the encoder feature maps. The transferred information helps the optimizer in optimizing the network more effectively \cite{Zhou2018}. \par
\item Segmentation blocks were also employed for self-ensembling and deep supervision purposes \cite{Isensee2018} within the decoder of the network. These segmentation blocks make predictions at each scale of the U-net by reducing the number of feature maps at each level of the decoder to the number of feature maps at the final output layer of the network. Combining these segmentation maps helps the network converge faster by transferring more information from the earlier levels of the decoder.
\end{enumerate}

The network was trained on both a multi-task framework as well as a cascaded binary segmentation task   to perform sequential segmentation of the tumour components,  and experiments were conducted to evaluate the performance of each model as well as an ensemble of both models. In the multi-task framework, the network output has three 3D channels of size 128\textsuperscript{3}, each of which corresponds to one of the three mutually inclusive tumour subregions and the network is trained to predict all tumour subregions at the same time. This multi-task learning regularizes the network by providing additional information related to the learning task. For the cascaded model, three cascaded networks were trained separately, one  for each of the three tumour subregions. Each of these networks has a single 3D output of size 128\textsuperscript{3} corresponding to one of the tumour subregions and at the time of inference, the input of each stage of the model is limited to patches containing the tumour region extracted by the previous cascaded stage. This means that the predicted tumour core is forced to lie inside the whole tumour, and the enhanced tumour core also  inside the tumour core region.  \par

\subsection{Training Procedure}
In the proposed method, all the four modalities were fed into the network at the same time to benefit from the information present in all.  Due to memory limitations of the Graphical Processing Unit (GPU) used, the network was trained with a batch size of 1.  To extract patches, each image was cropped so that the manually segmented tumour was located in the middle of each patch. Overlapping patches were extracted if the size of tumour was bigger than 128 voxels in any dimension. This strategy of extracting appropriate patches addresses the class imbalance issue to some extent, and also reduces the training time significantly. 
To further address the class imbalance problem, Dice Score was used as the loss function \cite{Shen2018b};  for the multi-task framework it was modified to take into account  the mean value of the Dice Scores of all output segmentation maps. Dice score (DS) and  multi-class Dice loss (MDS) function are expressed as:

\[
DS=2\times\frac{Y_{true} \times Y_{pred}}{Y_{true} + Y_{pred} + \epsilon} \;\;\;\;\;\;\;\;\;\;\; MDL = \frac{-1}{n}\sum_{1}^{n}DS
\]

where $Y_{true}$ and $Y_{pred}$ are the reference segmentation map (gold standard) and the predicted segmentation map for each of the output channels respectively. $n$ is the number of channels (3 for the multi-task model), and $\epsilon$ is a small value used to avoid division by zero (we set $\epsilon$ = 0.00001)\par
The network was trained using the Keras framework with TensorFlow backend, on an Nvidia Tesla Volta V100 GPU.  The GPU memory allowed  training of the network with batches of size 1 while allowing 16 filters in the highest level of the Dense U-net. The network was trained using Adam Optimizer with an initial learning rate of 5e-4, and reducing by a factor of 0.5 if the validation accuracy was not improving within the last 10 epochs. Dropout with a rate of 0.3 and also L2 norm regularization with a weight of 1e-5 were used for  regularization purposes. The network was programmed to be trained for a maximum of 300 epochs or until the validation accuracy did not improve in the last 50 epochs.

\subsection{Tumour localization}
To increase the prediction accuracy, a cascaded method was used in which the whole tumour region was localized using low-resolution images. To do so, the images were resized to 128\textsuperscript{3} voxels and then the model was applied to obtain an estimate of the tumour location. The prediction map was then resized to its original size and patches of size 128\textsuperscript{3} voxels were extracted such that the whole tumour is in the middle of that patch. If the predicted whole tumour is larger than 128\textsuperscript{3} voxels, more overlapping patches were extracted to cover the whole tumour region. The extracted patches were then used to predict the three tumour subregions. Finally, the segmentation maps were reconstructed by zero-padding, considering the location of the extracted patches. This method of patch extraction reduces both the test time and the rate of false positives prediction.

\subsection{Post-processing}
To further improve the accuracy of brain tumour segmentation, a post-processing method was applied. The post-processing may be summarized as follows:
\begin{enumerate}
\item  It was ensured that the hierarchy of the tumour subregions was respected. This means that the enhancing tumour subregion is inside the tumour core, and the tumour core is inside the whole tumour. To achieve this, any enhancing tumour voxels appearing outside the tumour core were removed, as well as any tumour core appearing outside the whole tumour. 
\item Connected components of any subregion smaller than 10 voxels were removed. 
\item Connected components of whole tumour that were primarily uncertain ( did not include any tumour core or enhancing tumour voxels, usually in cerebellum region, and/or having higher intensities in only FLAIR or T2 modalities). Such simple post-processing worked well in discarding false positive predictions possibly due to inherent system noise. This heuristics was verified by examining validation subject having such connected components as whole tumour. 
\item  Considering the fact that low grade gliomas (LGG) patients may have no enhancing tumour sub-region and also inspired by other work \cite{Isensee2018}, all enhancing tumour regions with less than 50 voxels were replaced by necrosis.
\end{enumerate}
After these steps, the remaining tumour subregions were fused to reconstruct the final segmentation map.

\section{Experiments and Results}\label{experimentsAndResults}
\begin{figure}
    \centering
    \includegraphics[width=\linewidth]{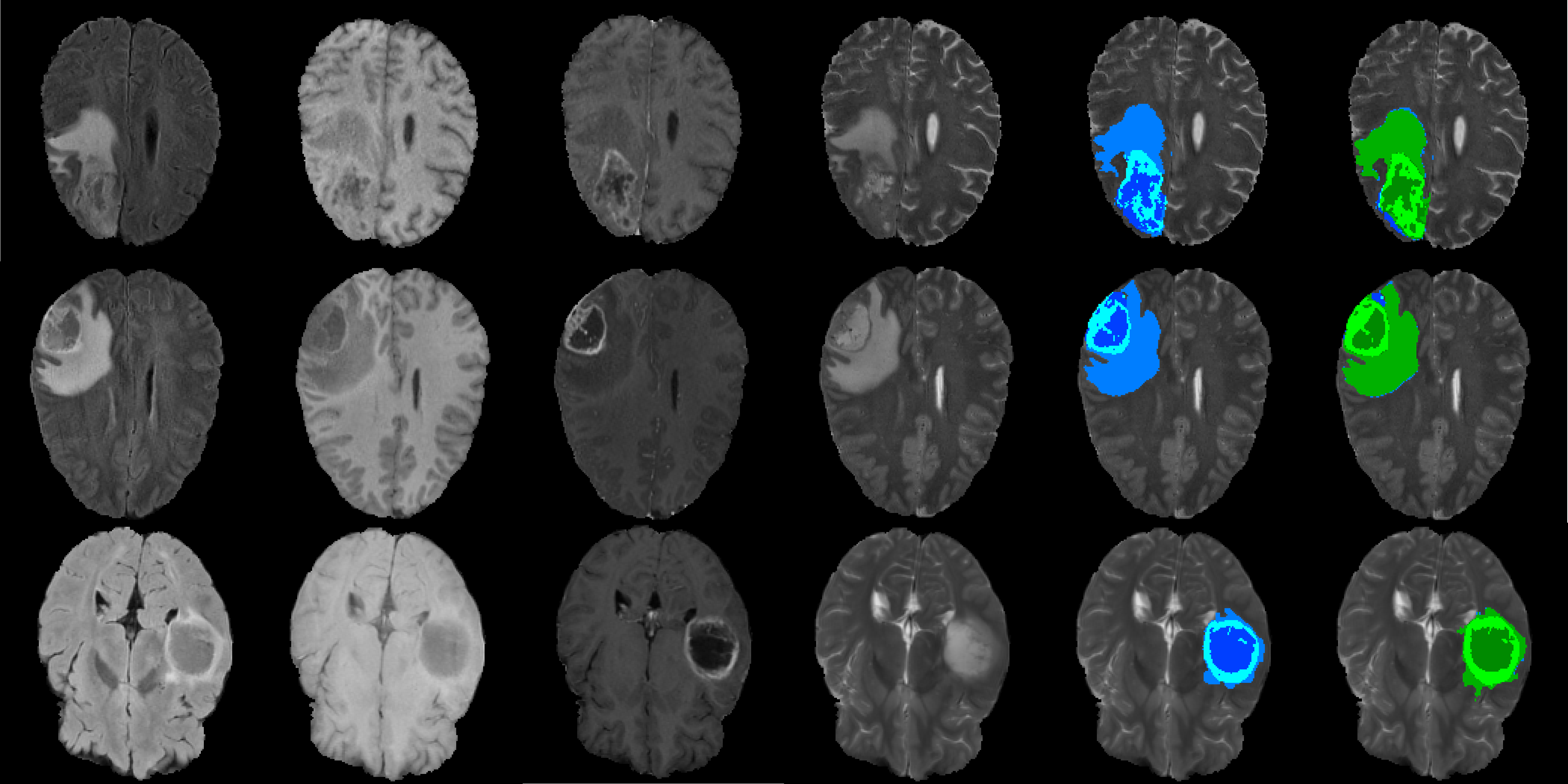}
    \caption{Qualitative results for three patients from BraTS20 training dataset. From left to right: Flair, T1, T1c, T2,  manual segmentation overlaid on T2, and model prediction.}\label{resultsVisulization}
\end{figure}

All scans in  the BraTS20 training dataset were used for training the proposed network in two different frameworks separately: cascaded binary segmentation task as well as the multi-task framework. Validation set results were obtained by using  each of these models separately as well as their ensemble. Ensembling was performed by averaging the Sigmoid probabilities of both model predictions. Thresholding and post-processing were then applied to the final prediction map. A threshold of 0.5 was chosen for whole tumour and tumour core subregion while for Enhancing tumour subregion the threshold was reduced to 0.4. This choice of threshold for enhancing tumour subregion was done based on the performance of the validation dataset and it enhanced the prediction accuracy for the subjects with under-segmentated prediction. However, it did not resolve the issue for some of the LGG cases as a result of which the Hausdorff distance is not as expected. So, in our ensemble experiment the enhancing core threshould was gradually decreased to 0.3, 0.2, and 0.1 if no enhancing tumour was not detected. This also resolved the under-segmentation issue for some of those LGG cases and enhanced the Hausdorff distance of enhancing tumour subregion (Table \ref{results}). Test time augmentation was applied for enhancing the prediction accuracy in all of the experiments. This was done by flipping the input image axes, and averaging the outputs of the resulting flipped segmentation probability maps. The results are listed in Table \ref{results}. All reported values were computed by the online evaluation platform. We also trained models using 5 fold cross-validation and a qualitative example generated using the trained model is depicted in Fig.~\ref{resultsVisulization}. Images include three sample cases from the training dataset, along with the manual annotations and  the model prediction. According to reported results, the multi-task model and the cascaded model have almost the same performance, while their ensemble resulted in a more accurate and robust model.

\begin{table}
    \begin{center}
\caption{Mean Dice Score (DSC)  and Hausdorff distance (95\%) (HD95), of the proposed segmentation method on BraTS 2020 validation dataset using the online IPP portal. ET: enhancing tumour, WT:whole tumour, TC: tumour core.}\label{results}
\scalebox{0.9}{
\resizebox{\columnwidth}{!}{
    \begin{tabular}{|l|c|c|c|c|c|c|}
        \hline
        Model &\multicolumn{3}{c|}{DSC } &\multicolumn{3}{c|}{DH95 } \\
        \hline
        Validation & ET & WT & TC & ET & WT & TC   \\
        \Xhline{3\arrayrulewidth}
        
        Multitask model & $0.76 \pm 0.28$ & $0.89 \pm 0.08$ & $0.82 \pm 0.17$  & $22.3 $ & $5.89 $ & $7.16 $ \\
        \hline
        
        Cascaded model  & $0.75  \pm 0.28$ & $0.88 \pm 0.13$ & $0.81 \pm 0.19$ & $17.17 $ & $7.16 $ & $10.4 $ \\
        \hline
        
        Ensemble model  & \boldsymbol{$0.78  \pm 0.26$}  & \boldsymbol{$0.90 \pm 0.06$} & \boldsymbol{$0.82 \pm 0.17$}  & \boldsymbol{$7.71 $} & \boldsymbol{$5.14$} & \boldsymbol{$6.64$} \\
        \hline
        
    \end{tabular}
}
}
\end{center}
\end{table}

\section{Conclusion}\label{conclusion}
In this paper a 3D densely connected U-net network was trained on the BraTS20 training dataset using two different approaches: a multi-task framework and a three-stage cascaded framework. The networks were trained using 3D patches of size 128\textsuperscript{3}. To address the class imbalance issue, instead of random patch selection, patches were selected by cropping images so that the manually segmented tumour was located in the middle of each patch. Brain tumour segmentation was performed in two stages: first a coarse prediction was performed on low resolution images to localize the tumour, and then fine segmentation was performed by extracting patches in the tumour region, applying the model to those patches, and finally stitching patches together. The model predictions were then filtered through a post-processing step in which connected components analysis was used to reduce the false positive predictions.
The performance of the multi-task model, the cascaded model and also an ensemble of both models was evaluated using the BraTS20 validation dataset, and results were calculated on the online CBICA Image Processing Portal. The reported average Dice Score for the ensemble model were 0.90, 0.82, and 0.78 for whole tumour, tumour core and enhancing tumour respectively. We suffered from some hardware limitations while training our models, which we believe has  impacted the model performance. For example, the maximum duration time for the cloud GPU we were using was 48 hours after which the training was terminated, and also due to GPU memory limitation we could only train with a batch size of one. We believe that with access to more powerful hardware,  the model performance would be further enhanced.

\bibliographystyle{splncs04}
\bibliography{referenceList}

\end{document}